\documentclass[aps,prb,reprint,showpacs,superscriptaddress,
footinbib,citeautoscript]{revtex4-1}

\usepackage{graphicx}

\usepackage[utf8]{inputenc}
\usepackage{csquotes}
\usepackage[american]{babel}
\usepackage[T1]{fontenc}
\usepackage{enumerate}
\usepackage{mdwlist}
\usepackage[activate=normal]{pdfcprot}
\usepackage{bbding}
\usepackage{color}
\usepackage{amssymb}
\usepackage{amsmath}
\usepackage{amsfonts}
\usepackage{mathrsfs}

\usepackage{bm}
\usepackage{dcolumn}
\usepackage{color}
\usepackage[colorlinks=true,citecolor=blue]{hyperref}
\hypersetup{colorlinks=true,citecolor=blue,linkcolor=red,urlcolor=blue}

\newcommand{\comment}[1]{}

\begin{document}

\title{Majorana Zero Modes in Superconducting Proximity-coupled Magnetic Domain Wall}
\date{\today}

\author{Babak Zare Rameshti}
\email{b.zare.r@ipm.ir}
\affiliation{School of Physics, Institute for Research in Fundamental Sciences (IPM), Tehran 19395-5531, Iran}
\author{Yaser Hajati}
\affiliation{Department of Physics, Shahid Chamran University of Ahvaz, Ahaz, Iran}
\author{Imam Makhfudz}
\affiliation{Laboratoire de Physique Théorique–IRSAMC, CNRS and Université de Toulouse, UPS, F-31062 Toulouse, France}

\begin{abstract}
We propose a simple model consisting of a magnetic domain wall proximity-coupled to an $s$-wave superconductor for realization of Majorana zero-energy modes. A spin-dependent gauge transformation translates the rotating magnetic profile through the domain wall to effective spin-orbit and Zeeman terms. The Hamiltonian breaks time reversal and chiral symmetries, while preserving particle-hole symmetry, placing itself into topological D class characterized by the $\mathbb{Z}_{2}$ topological invariant for quasi one-dimensional system. 
The low-energy sector of the model maps to the one isomorphic with Kitaev Hamiltonian. The existence and localization of Majorana zero modes in the nontrivial phase are demonstrated explicitly and we obtain the topological phase diagram with extended regime of nontrivial phase and surprising occurrence of a re-entrance phase transition. Our calculation shows that the system can be easily tuned between trivial and topological ground states and can be implemented experimentally to realize non-Abelian statistics.
\end{abstract}

\maketitle
\textit{Introduction.\textemdash}\label{intro}
Majorana fermions (MFs) are quantum particles which are their own anti-particles\cite{Wilczek, Kitaev, FuKane, MZHKane, Beenakker1, Franz}. Because of their topological properties and the potential applications in fault-tolerant topological quantum computations, the search for MFs in solid state systems is thus of big general interest\cite{Levi, Service}. Topological quantum computation is manipulation of the wave function within a degenerate many-body ground state of many non-Abelian anyons and such topological quantum computation, in contrast to ordinary quantum computation, would not require any quantum error correction since the Majorana excitations are naturally immune to errors \cite{Kitaev}. Many solid state materials have been predicted to be candidates for realization of MFs\cite{Sato, Diehl, Jiang, Sau, lutchyn}. Even though experimental progress in the solid state systems has been made in the past few years, an evidence of MFs is still lacking due to many factors influencing the measurement results in solid state materials. 

The first system that was predicted to show MFs consisted of a quasi one-dimension semiconducting wire with strong spin-orbit interactions in proximity to an $s$-wave superconductor and with an applied Zeeman magnetic field\cite{Rafeal} which effectively reduces to Kitaev's model, known to support Majorana modes at its end\cite{Kitaev}. There have been several interesting proposals to realize MFs in one-dimensional systems with an effective rather than intrinsic spin-orbit coupling\cite{Klinovaja, Choy, Kjaergaard, Perge}.
\comment{Braunecker et al.\cite{Braunecker, Gangadharaiah, Braunecker2} showed that in one-dimensional conductors, an external magnetic field  is equivalent to a Rashba spin-orbit interaction in the local frame. This situation can be seen in a quantum wire with nuclear spin ordering. Choy et al.\cite{Choy} proposed an alternative system consisting of a one-dimensional chain of magnetic nanoparticles on a superconductor to create a similar situation. Kjaergaad et al.\cite{Kjaergaard} showed that a nanowire in proximity with an s-wave superconductor in an external magnetic field that rotates along the wire, is equivalent to a nanowire with Rashba-type spin-orbit coupling. Also, Nadj-Perge et al.\cite{Perge} demonstrated that a chain of magnetic atoms on surface of an s-wave superconductor can host MF modes. In this system a non-collinear arrangement of magnetic moments of adjacent atoms is essential to realize robust MF end modes.}The search for MFs in these systems is both rewarding and difficult as a definite experimental observation of the predicted MFs has so far remained formidable.
\begin{figure}
\begin{center}
\includegraphics[scale=0.08]{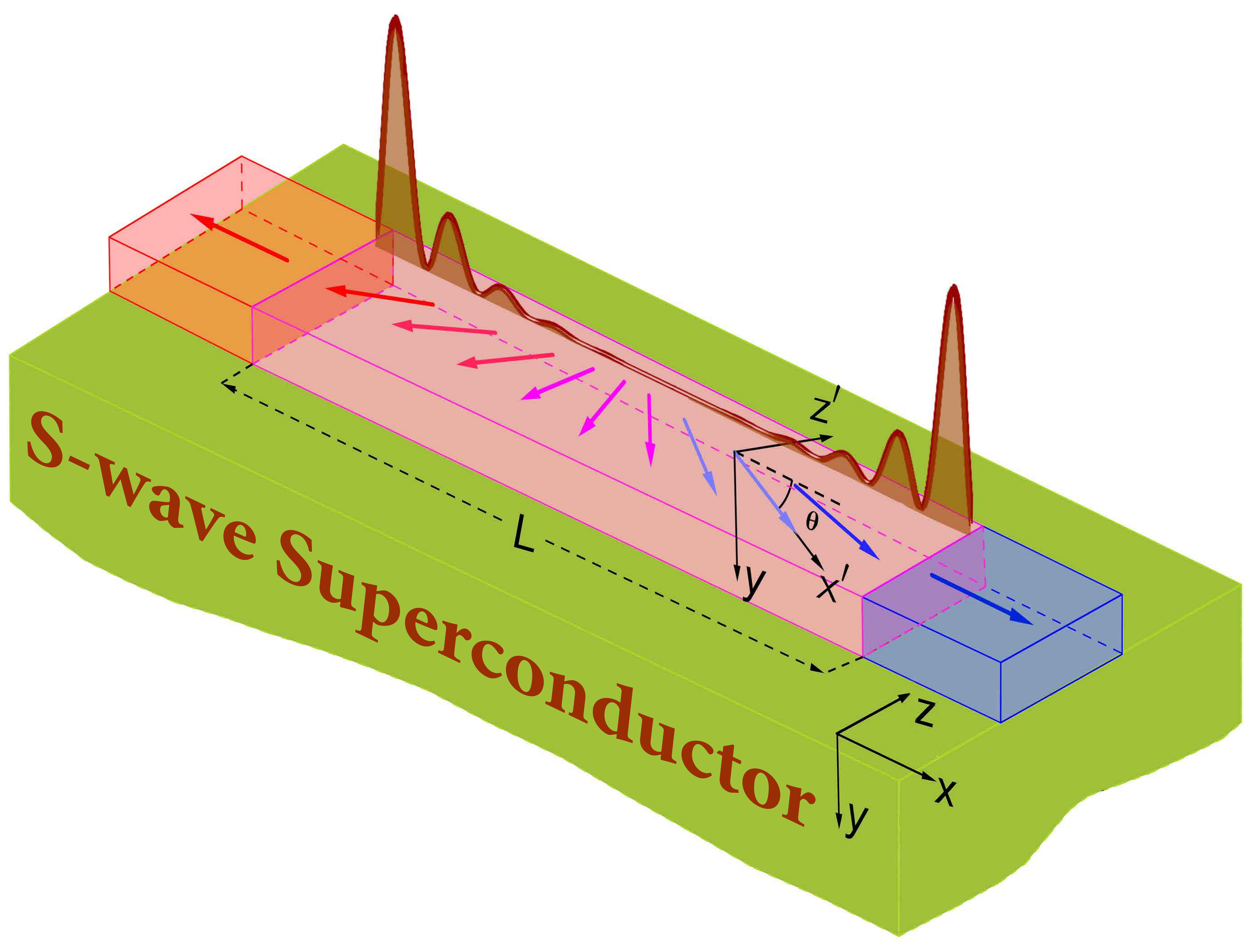}
\end{center}
\caption{(Color online) A schematic configuration of proposed device for realizing Majorana fermions in one-dimensional magnetic domain wall, with the length of $L$, proximity coupled to a conventional $s$-wave superconductor.}\label{fig1}
\end{figure}
\par
Here, we describe a very simple model to realize MFs using a ferromagnetic domain wall (DW) on top of an s-wave superconductor (as Fig. \ref{fig1}), without requiring intrinsic spin-orbit coupling and an external magnetic field. Domain wall is a structure with spatially-varying magnetization configuration which separates regions with uniform magnetization. It is widely studied and also easily fabricated in magnets of submicrometer size\cite{Kent, Garcia, Parkin, Tatara, Tatara2, Chopra, Parkin2, Linder, Linder2}. Electronic transport through ferromagnetic domain walls has been currently a subject of extensive investigations, both theoretically and experimentally\cite{Tatara3, Yamaguchi, Han, Wang, Shiba, Lei, Kagawa, Chen, Berger, Garcia, Tatara, Chopra}. Now, such a domain wall on an s-wave superconductor provides the necessary ingredient to generate MFs. It will be shown here that by putting a DW on top of an $s$-wave superconductor, MFs will be present at the ends of the DW.
\par

\textit{Model.\textemdash}\label{sec:model}
We propose a setup for realizing Majorana fermions in one-dimensional magnetic domain wall, with length $L$, in proximity with an isotropic $s$-wave superconductor as shown in Fig. (\ref{fig1}). The magnetic domain wall is sandwiched between two ferromagnets with uniform magnetizations in opposite directions along $x$ axis in the lab frame. The electronic properties of the DW are modeled by two-band $s$-$d$ Stoner model, described by an effective Hamiltonian of the form
\begin{equation}
\mathcal{H}_{0}=\frac{p^2}{2m^{*}}-\mu-\textbf{V}_{\rm ex}(\textbf{r})\cdot\bm{\sigma}
\end{equation}
where $\bm{\sigma}=(\sigma_{x}, \sigma_{y}, \sigma_{z})$ is the vector of Pauli spin matrices and $m^*$ and $\mu$ are the effective mass and chemical potential, respectively. The last term represents the exchange coupling between the $s$ conduction electron spin and the $d$ electron spin of the local magnetization $\textbf{V}_{\rm ex}=V_{\rm ex}\bm{n}(\textbf{r})$, where $V_{\rm ex}$ is the spin splitting strength and $\bm{n}(\textbf{r})$ the direction of the local magnetization. 

The functional form of the $\bm{V}_{\rm ex}(\textbf{r})$ describes the shape of magnetic domain wall. We consider a "N\'{e}el wall" with magnetization vector parallel to the $x$ axis in the leads far from the domain wall center and turns by $180^\mathrm{o}$ in the $xz$ plane within the wall. We assume a trigonometric magnetization profile in the DW, $\bm{V}_{\rm ex}=(\cos\theta(x), 0, \sin\theta(x))$ where $\theta(x)=\nu\pi x/L$ with $\nu$ being the winding number, the total angle (phase) change along the system, and then do a spin-dependent gauge transformation in spin space from the fixed reference frame to the rotated frame, which is in the direction of local magnetization vector $\bm{V}_{\rm ex}(\textbf{r})$. In our model, it is given by a rotation about the $y$ axis $\mathcal{R}=e^{i\theta\sigma_{y}/2}$. The DW Hamiltonian in the rotating frame $\mathcal{H}_{\mathrm{r}}$ takes the form
\begin{eqnarray}
\mathcal{H}_{\mathrm{r}}&=&\mathcal{R}^{-1}_{y}(\theta)\mathcal{H}_{0}\mathcal{R}_{y}(\theta)=e^{-i\theta\sigma_{y}/2}\mathcal{H}_{0}e^{i\theta\sigma_{y}/2}\nonumber\\
&=&\frac{p^2}{2m^{*}}-\mu+\frac{\hbar^{2}\omega^{2}}{2m^{*}}-\frac{\hbar}{m^{*}}\omega p\sigma_{y}+V_{\rm ex}\sigma_{x}\label{eq2}
\end{eqnarray}
where $\omega=\partial_{x}\theta(x)/2$. Due to the gauge transformation the electrons experience an effective spin-orbit interaction $(\hbar\omega/m^{*})p\sigma_{y}$, Zeeman field $V_{\rm ex}\sigma_{x}$ and modified chemical potential $\mu-\hbar^{2}\omega^{2}/2m^{*}$ in the rotated frame. The wave function in the fixed reference frame (along the $x$ axis) can be obtained from the relation $\Psi_{\sigma}(x)=\mathcal{R}\Phi_{\sigma}(x)$. 

By inducing superconductivity into the DW via proximity to a conventional $s$-wave superconductor described within the effective BCS mean-field approximation, the low-energy effective single-particle BdG Hamiltonian in the basis $\Phi(x)=(u_{\uparrow}(x), u_{\downarrow}(x), v_{\downarrow}(x), -v_{\uparrow}(x))$ can be written as
\begin{eqnarray}
\mathcal{H}_{\mathrm{BdG}}&=&-\left(\frac{\hbar^{2}}{2m^{*}}\partial^{2}_{x}+\mu^{\prime}\right)\tau_{z}\otimes\sigma_{0}+i\frac{\hbar^{2}}{m^{*}}\omega \partial_{x}\tau_{z}\otimes\sigma_{y}\nonumber~~\\
&&+V_{\rm ex}\tau_{0}\otimes\sigma_{x}+\Delta_{0}\tau_{x}\otimes\sigma_{0}\label{eq3}
\end{eqnarray}
where $\mu^{\prime}=\mu-\hbar^{2}\omega^{2}/2m^{*}$ and the $\tau_{i}$ Pauli matrices act in the particle-hole (Nambu) space. It can
easily be verified that the BdG Hamiltonian has a particle-hole symmetry (PHS) $\Xi\mathcal{H}_{\mathrm{BdG}}\Xi^{-1}=-\mathcal{H}_{\mathrm{BdG}}$ with respect to the \textit{particle-hole} anti-unitary operator $\Xi=\tau_{y}\sigma_{y}\mathcal{K}$, where $\mathcal{K}$ is the complex conjugate operator, satisfying $\Xi\Phi(x)=\lambda\Phi(x)$. Both the time-reversal symmetry (TRS), represented by the operator $\Theta=i\sigma_{y}\mathcal{K}$, and the chiral symmetry, represented by the operator $\mathcal{C}=-i\tau_{y}\mathcal{K}$ are broken. The absence of TRS and chiral symmetry in the presence of PHS ensures that the Hamiltonian is in the D symmetry class characterized by the $\mathbb{Z}_{2}$ topological invariant for the quasi one-dimensional system\cite{Altland}. This implies that, under appropriate conditions, the system can support localized Majorana modes that remain topologically protected against local perturbations.
\par
\begin{figure}
\includegraphics[width=8.0cm]{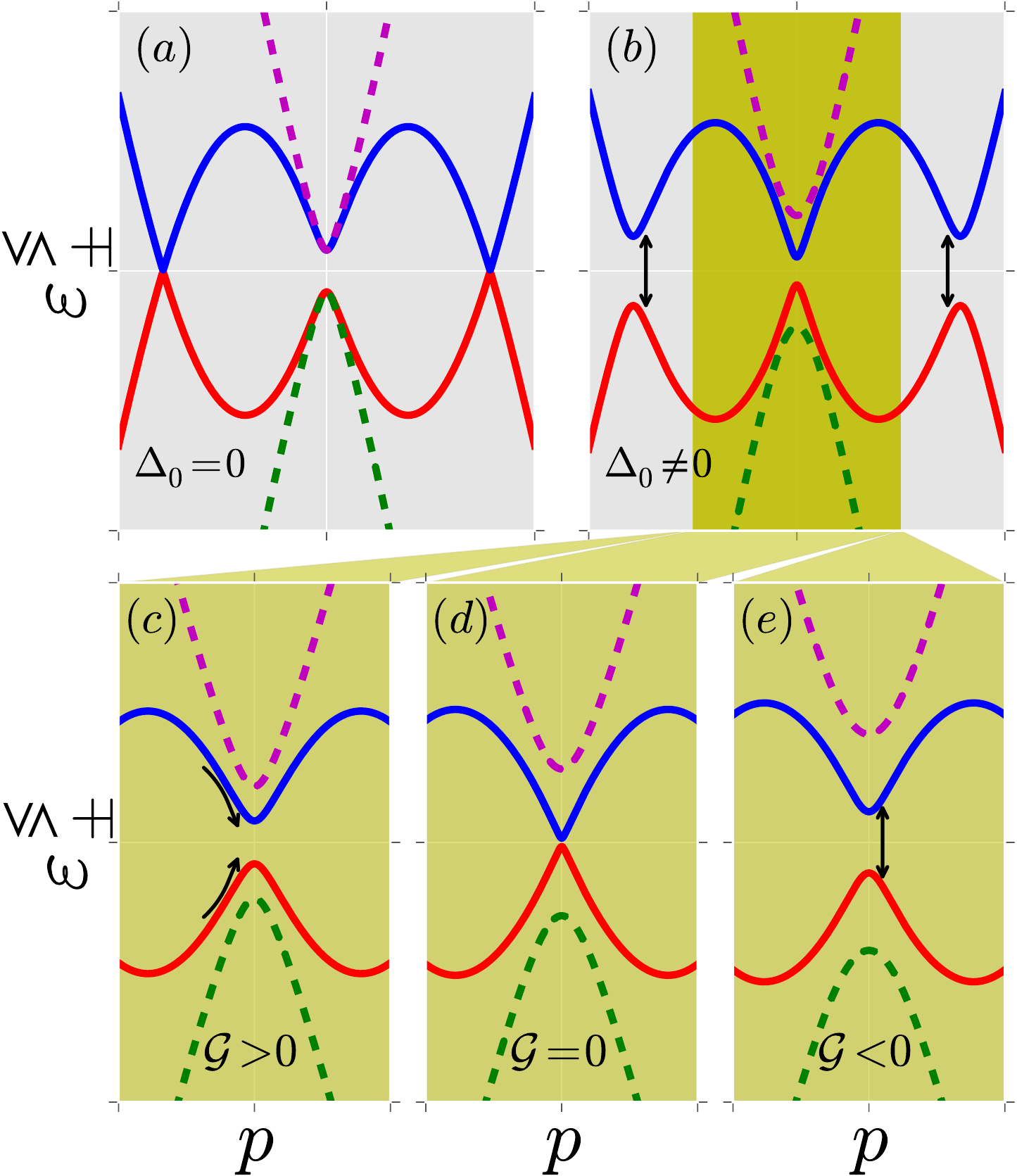}
\caption{(Color online) The bulk spectrum of the BdG Hamiltonian is shown where the blue (red) line is for $\varepsilon^{>}_-(p)$ ($\varepsilon^{<}_-(p)$), while dashed lines are the corresponding high energy excitations $\varepsilon^{>,<}_+(p)$. (a) The spectrum with $\Delta_{0}=0$ contains an energy gap at the $p=0$ point. (b) The $\Delta_{0}$ opens a gap at two degenerate points. The topological quantum phase transition characterized by the closing and reopening the gap $\varepsilon_{\mathrm{gap}}$ (bottom panels).}\label{fig2}
\end{figure}

The spectrum of the bulk states is given by
\begin{eqnarray}
\varepsilon^{2}_{\pm}(p)&=&\xi_{p}^{2}+V_{\rm ex}^{2}+\Delta^{2}_{0}+(\hbar\omega p/m^{*})^{2}\nonumber\\
&&\pm 2\sqrt{V_{\rm ex}^{2}(\Delta_{0}^{2}+\xi_{p}^{2})+(\hbar\omega p\xi_{p}/m^{*})^{2}}
\end{eqnarray}
where $\xi_{p}=p^{2}/2m^{*}-\mu^{\prime}$ and the two $\pm$ branches are due to the spin splitting as shown in Fig.(\ref{fig2}). The minimum gap is between the two excitation branches $\varepsilon^{>,<}_{-}(p)$ at $p=0$ (measured from Fermi level) given by
\begin{eqnarray}
\varepsilon_{\mathrm{gap}}=\frac{1}{2}\left\vert\varepsilon^{>}_{-}(0)-\varepsilon^{<}_{-}(0)\right\vert=\left\vert V_{\rm ex}-\sqrt{\Delta^{2}_{0}+\mu^{\prime 2}}\right\vert
\end{eqnarray}
which can be closed at the critical point $ V_{\rm ex}^{2}=\Delta^{2}_{0}+\mu^{\prime 2}$, where the superscript $>(<)$ refers to the energy band above (below) the Fermi level ($E=0$). The spectrum of the magnetic domain wall comprises two spin-splitted bands, along with two non-gapped degenerate Kramer's pairs and the energy gap at the $p=0$ point due to the broken time-reversal symmetry, for the electrons and their particle-hole related partners with the negative energy as indicated in Fig.\ref{fig2}(a). Inducing the isotropic superconducting pair potential $\Delta_{0}$ via the proximity with an $s$-wave conventional superconductor opens a gap at the outer wings of the dispersion, which eliminates the possibility of high-momentum gapless excitations and leaving only the chiral states near $p=0$ as the low energy excitations and modifies the gap forming near $p=0$, as demonstrated in Fig.\ref{fig2}(b). The topological quantum phase transition, specified by the closing and reopening of the gap $\varepsilon_{\mathrm{gap}}$ with opposite sign is shown in the bottom panels of Fig.(\ref{fig2}), where the system goes from the superconducting-dominated trivial gap Fig.\ref{fig2}(c), to the exchange field-dominated nontrivial gap Fig.\ref{fig2}(e) while pass the critical point Fig.\ref{fig2}(d)\cite{Choy}. 
\par
While the full BdG Hamiltonian Eq. (\ref{eq3}) is not that of a spinless
$p$-wave superconductor, if the Fermi energy lies between the two bands $\varepsilon^{>,<}_{-}(p)$, the system is effectively an one-dimensional spinless system. The effective low energy physics of the BdG Hamiltonian Eq. (\ref{eq3}) then exactly maps to the one-dimensional spinless $p$-wave superconductor. To this end, we consider the effective Hamiltonian Eq. (\ref{eq2}) without superconducting pair potential $\Delta_{0}=0$ and diagonalize it using a unitary transformation, $\mathcal{H}_{0}=\mathcal{U}\Lambda \mathcal{U}^{\dagger}$ where $\Lambda=\mathrm{diag}(\varepsilon_{+}, \varepsilon_{-}, -\varepsilon_{-}, -\varepsilon_{+})$ with the eigenvalues given by,
\begin{equation}
\varepsilon_{\pm}=p^{2}/2m^{*}-\mu^{\prime}\pm\sqrt{V_{\rm ex}^{2}+(\hbar\omega p/m^{*})^{2}}
\end{equation}
We then project the BdG Hamiltonian onto the Hilbert space of the eigenstates of the Hamiltonian $\mathcal{H}_{0}$ without superconductivity using the same unitary transformation, $\mathcal{H}_{\mathrm{tr}}=\mathcal{U}^{\dagger}\mathcal{H}_{\mathrm{BdG}}\mathcal{U}$, which yields
\begin{equation}
\mathcal{H}_{\mathrm{tr}}=\Lambda+\Delta_{0}\cos\vartheta\tau_{x}-\Delta_{0}\sin\vartheta\tau_{y}\sigma_{x}
\end{equation}
where the angle $\vartheta$ is given by the relation $\sqrt{V_{\rm ex}^{2}+(\hbar\omega p/m^{*})^{2}}\sin\vartheta =\hbar\omega p/m^{*}$. 
To study the Majorana zero modes of interest, it suffices to retain the $\varepsilon_{-}(p)$ branch of $\mathcal{H}_{\mathrm{tr}}$ to catch the low energy physics, which gives an effective Hamiltonian,
\begin{eqnarray}
\mathcal{H}_{\mathrm{pr}}= \sum_{p}\left(\varepsilon_{-}(p)c^{\dagger}_{p}c_{p}+\tilde{\Delta}(p)c^{\dagger}_{p}c^{\dagger}_{-p}+\mathrm{h.c.}\right)
\end{eqnarray}
where 
\begin{equation}\label{effectivepairgap}
\tilde{\Delta}(p)=\Delta_{0}\frac{(\hbar\omega p/m^{*})}{[V_{\rm ex}^{2}+(\hbar\omega p/m^{*})^{2}]^{1/2}}
\end{equation}
is an effective superconducting
pair potential with the desired $p$-wave symmetry. The above Hamiltonian is isomorphic to the Kitaev Hamiltonian \cite{Kitaev}, and is thus expected to host localized Majorana states bound to defects in the superconductor, where the effective order parameter amplitude vanishes.
\par
\textit{Majorana Zero Modes Solution.\textemdash}
The full BdG Hamiltonian Eq. (\ref{eq3}) can be diagonalized by using a Bogoliubov transformation with the quasiparticle operators $\alpha$ and $\alpha^{\dagger}$ satisfying the anti-commutation relations for complex fermions. The corresponding self-adjoint Majorana operators $\gamma_{+}=(\alpha^{\dagger}+\alpha)/2$ and $\gamma_{-}=i(\alpha^{\dagger}-\alpha)/2$ are given by $\gamma_{+}=\int dx [u_{\uparrow, +}(x)\Phi_{\uparrow}^{\dagger}(x)+u_{\downarrow, +}(x)\Phi_{\downarrow}^{\dagger}(x)+ u_{\uparrow, +}(x)\Phi_{\uparrow}(x)+u_{\downarrow, +}(x)\Phi_{\downarrow}(x)]$,
$\gamma_{-}=i\int dx [u_{\uparrow, -}(x)\Phi_{\uparrow}^{\dagger}(x)+u_{\downarrow, -}(x)\Phi_{\downarrow}^{\dagger}(x)+u_{\uparrow, -}(x)\Phi_{\uparrow}(x)-u_{\downarrow, -}(x)\Phi_{\downarrow}(x)]$
where $u_{\sigma,\pm}(x)$, $v_{\sigma,\pm}(x)$ are the solutions to the BdG equations subject to boundary conditions and $\pm$ denotes the solutions for the positive and negative energy bands, respectively. It can easily be demonstrated from the explicit expressions for the Majorana operators given above that $\gamma_{\pm}=\gamma_{\pm}^{\dagger}$. In our particle-hole symmetric system, these Majorana fermions can only occur from zero-energy states.  
\begin{figure}
\includegraphics[width=7.50cm]{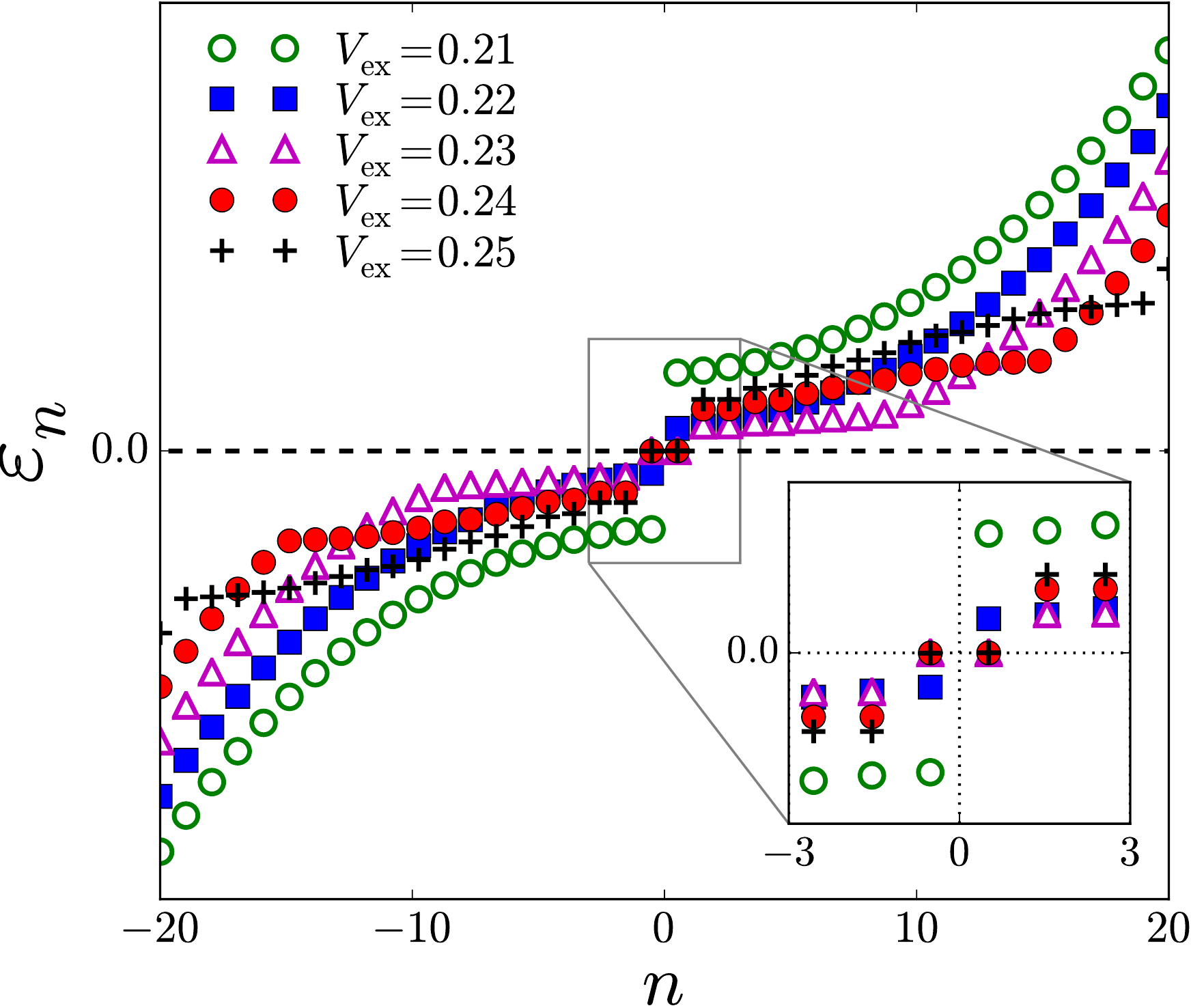}
\caption{(Color online) The quasi-particle excitations spectra where the mode number $n$ labels the eigenvalues of the BdG Hamiltonian. We have used $\tilde{\mu}=-0.1$ and $\tilde{\Delta}_{0}=0.2$ in normalized units where $E=(\hbar^2\bar{\omega}^2)/m^*=1$ with $\bar{\omega}=\pi/(2L)$. Inset: the midgap states in a nontrivial topological phase with Majorana fermions.}\label{fig3} 
\end{figure}
\par
\begin{figure}
\includegraphics[width=8.6cm]{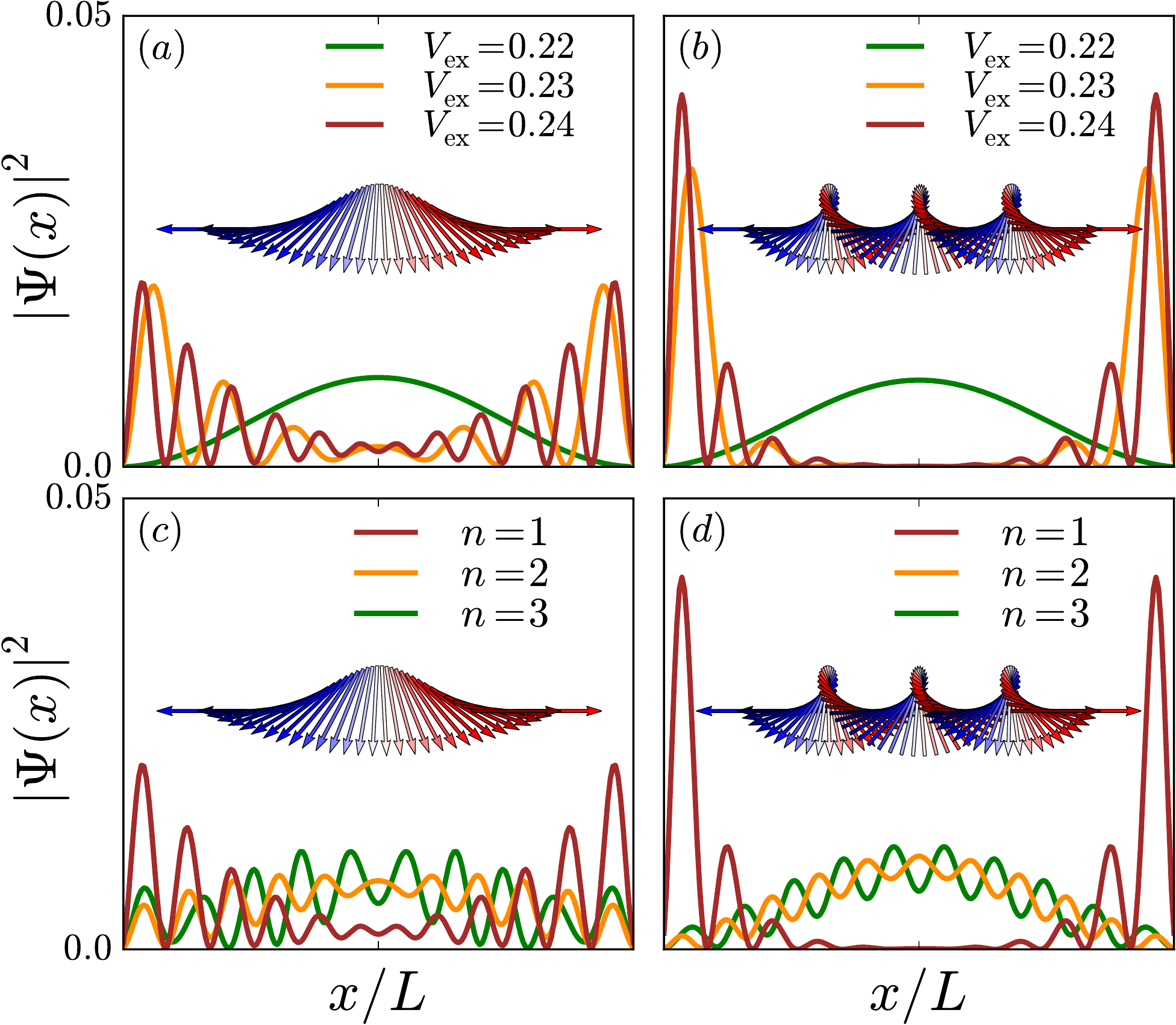}
\caption{(Color online) The wave function amplitudes of the lowest energy states with the same parameter set as in Fig. (\ref{fig3}), for the mode $n=1$ versus $V_{\mathrm{ex}}$ ((a) and (b)) and versus $n$, showing localized Majorana state ($n=1$) and extended higher modes ((c) and (d)).}\label{fig4}
\end{figure}

The low-energy BdG excitations spectrum $\varepsilon_{n}$ and their hole equivalent of our proposal are shown in Fig. (\ref{fig3}). The induced superconducting gap gets suppressed by the exchange field for the lowest energy modes in the trivial phase and eventually changes sign by transition to the nontrivial phase where one clearly sees the existence of zero-energy modes; this is the Majorana state we are looking for. To show the localized nature of this state, the corresponding wave function amplitudes for the lowest mode $n=1$ has been calculated in both side of the critical point as indicated in Fig. \ref{fig4}(a)-(b), showing extended (localized) wave functions in the trivial (nontrivial) state. In Fig. \ref{fig4}(c)-(d) the amplitude of the wave functions has been plotted for different modes $n$,  showing localized Majorana state ($n=1$) and extended higher modes. The Majorana wave function envelope decays exponentially with distance into the the bulk from the boundaries with localization length determined by the effective superconducting coherence length $\xi\sim\hbar v_F/\tilde{\Delta}$. The effective pairing gap $\tilde{\Delta}$ can be obtained by linearizing $\tilde{\Delta}(p)$ in Eq.(\ref{effectivepairgap}) around $p=0$ which gives $\tilde{\Delta}=\hbar v_F(\omega\Delta_0)/V_{\mathrm{ex}}$ and hence $\xi\sim V_{\mathrm{ex}}/(\omega\Delta_0)$ where $\omega=\pi\nu/(2L)$. The overlap of the two Majorana states at both ends is proportional to $e^{-L/\xi}$, hence increasing the winding number $\nu$ makes the end states to be more independent as indicated in Figs. \ref{fig4}(b)-(d).
\par
To gain deeper insight into the Majorana zero-energy modes, we solve the BdG equation analytically with $\varepsilon_1=0$. Using the particle-hole symmetry, the BdG equations can be reduced to dimensionless $2\times2$ matrix form as
\begin{equation}
\begin{pmatrix}
-\frac{1}{2}\partial^{2}_{\tilde{x}}-\tilde{\mu}^{\prime}+\tilde{V}_{\rm ex} & \nu\partial_{\tilde{x}}+\lambda\tilde{\Delta}_{0}\\
-\nu\partial_{\tilde{x}}-\lambda\tilde{\Delta}_{0} & -\frac{1}{2}\partial^{2}_{\tilde{x}}-\tilde{\mu}^{\prime}-\tilde{V}_{\rm ex}
\end{pmatrix}\begin{pmatrix}
u_{\uparrow}(\tilde{x})\\
u_{\downarrow}(\tilde{x})
\end{pmatrix}=0
\end{equation}
Decoupling this system of coupled second order differential equations yields a fourth order homogeneous differential equation for each $u_{\sigma}(\tilde{x})$, which in general has solutions of the form $u_{\sigma}(\tilde{x})\sim e^{m\tilde{x}}$ and the corresponding characteristic equation for $m$ reads,
\begin{equation}
m^{4}+4(\tilde{\mu}^{\prime}+\nu^{2})m^{2}+8\nu\lambda\tilde{\Delta}_{0}m+4\mathcal{G}=0
\end{equation}
where $\mathcal{G}=\tilde{\Delta}_{0}^{2}+\tilde{\mu}^{\prime2}-\tilde{V}_{\rm ex}^{2}$. The roots of the polynomial should satisfy the following constraints,
\begin{equation}
\prod_{n=1}^{4}m_{n}=4\mathcal{G},\qquad\sum_{n=1}^{4}m_{n}=0.\label{eq19}
\end{equation}
It can be verified using Eq. (\ref{eq19}) that for $\mathcal{G}>0$ ($\mathcal{G}<0$), there are always an even (odd) number of solutions with positive real part $\text{Re}(m)>0$, implying an even (odd) number of band crossings at the Fermi level. The boundary and normalizability conditions for a localized Majorana wave function solution can only be satisfied when $\mathcal{G}<0$, permitting the zero-energy modes to exist. We can thus conclude that $\mathcal{G}=0$ determines the topological quantum phase transition indicated by the closing and reopening the bulk gap at the topological critical point.
\par
\begin{figure}
\includegraphics[width=8.0cm]{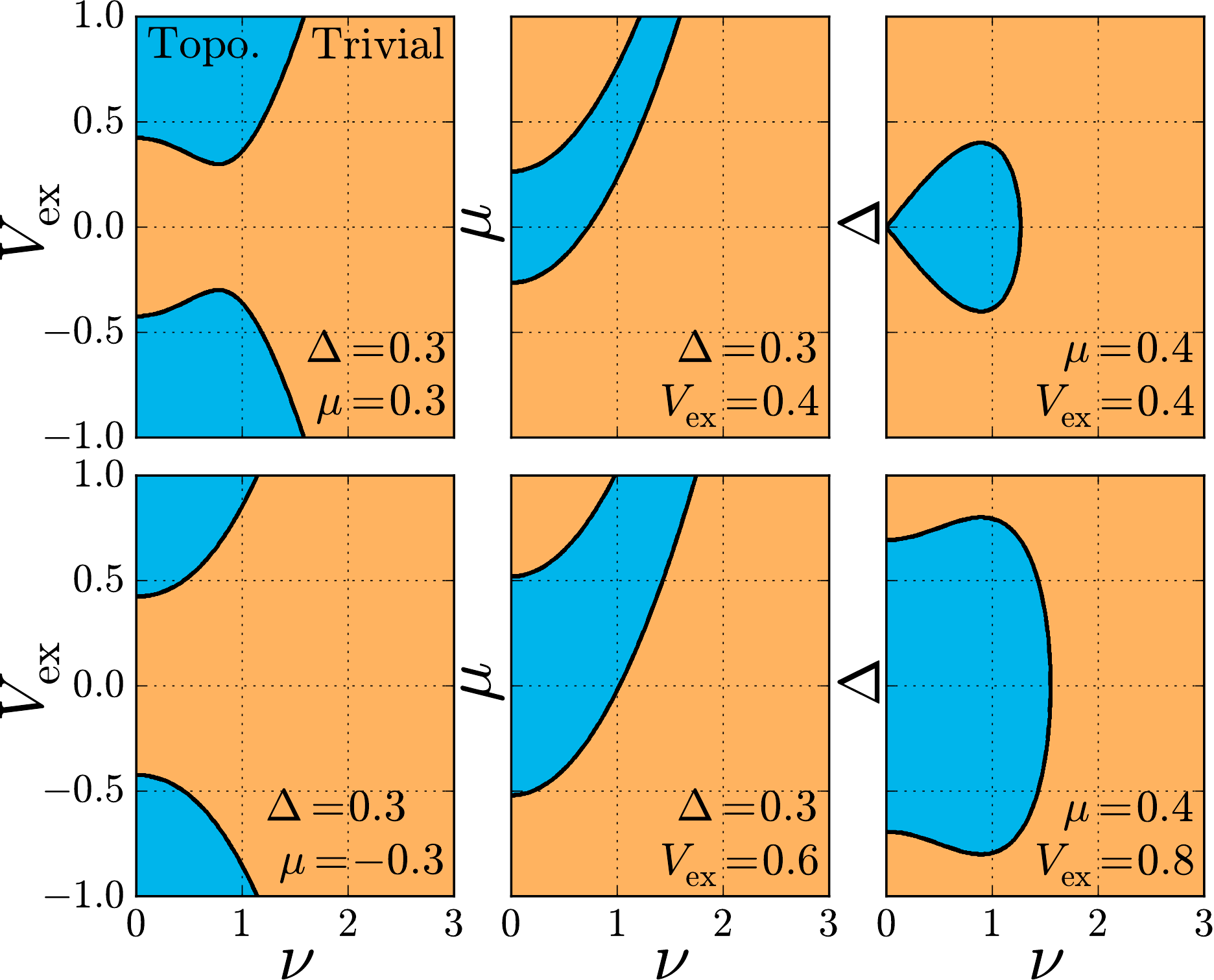}
\caption{(Color online) The topological phase diagrams of the model as functions of microscopic parameters $V_{\mathrm{ex}}$, $\Delta_0$, $\mu$, and $\nu$ where the light blue (light brown) area corresponds to the topologically nontrivial (trivial) phase with $\mathcal{Q}=-1(+1)$. The phase boundary (black lines) correspond to $\mathcal{G}=0$.}\label{fig5}
\end{figure}

The presence of the zero-energy self-adjoint eigenstate thus relates directly to the sign of the gap $\varepsilon_{\mathrm{gap}}$ and thus that of $\mathcal{G}$. The sign of $\mathcal{G}$ then characterizes the topological nature of the system and defines the associated topological invariant, which is a $\mathbb{Z}_2$ number. 
We compute the $\mathbb{Z}_2$ topological invariant $\mathcal{Q}$ of our model, given by $\mathcal{Q}=\mathrm{sign}(\mathcal{G})=\mathrm{sign}(\Pi_n m_n)=\mathrm{Pf}(\mathcal{H}_{{\rm BdG}})$, where $\mathrm{Pf}$ refers to the Pfaffian of the BdG Hamiltonian. We then deduce the topological phase diagram in the space of microscopic parameters $V_{\mathrm{ex}}$, $\Delta_0$, $\mu$, and $\nu$. The results are shown in Fig. (\ref{fig5}). We note that the topologically nontrivial phase exists within extended portion of the phase diagram. Surprisingly, re-entrance phase transition occurs in the phase diagram, where by varying appropriate parameters, one can go from trivial to nontrivial and back to trivial phase or vice versa. To have actually Majorana modes, one should have both $\mathcal{Q}=-1$ and gapped bulk states. For the ferromagnetic limit $\nu=0$, in the whole range where $\mathcal{Q}=-1$ bulk is gapless so it does not have topological phase while in the antiferromagnetic limit $\nu=\infty$ the $\mathcal{Q}$ never changes sign and we thus never cross the phase transition.
\par
A stability analysis generalizing that in \cite{Alicea} suggests that in order for the Majorana bound states to be stable, we have to choose the parameters to be such that $0<|V_{\mathrm{ex}}-\sqrt{\Delta^2_0+\mu'^2}|< \Delta_0$. This ensures that the central gap (at $p=0$) is nonzero and at the same time smaller than the two outer gaps which, in the limit of $V_{\mathrm{ex}}\ll \hbar\omega p_F/m^*$ where $\pm p_F$ are the locations of the two outer gaps, is given by $\Delta(p_F)\approx\Delta_0$. Considering the regime of dominant exchange field energy, which favors the nontrivial topological phase hosting MFs, the nonzero gap condition of the above criterion translates into $|\mu'|<\sqrt{V^2_{\mathrm{ex}}-\Delta^2_0}$ or equivalently $|V_{\mathrm{ex}}|>\sqrt{\Delta^2_0+\mu'^2}$ or $|\Delta_0|<\sqrt{V^2_{\mathrm{ex}}-\mu'^2}$ where $\mu'=\mu-(\hbar^2\pi^2\nu^2)/(8m^*L^2)$. It can be checked that these criteria agree with the phase diagram Fig.(\ref{fig5}). The braiding of Majorana fermions to realize a non-Abelian statistics can be implemented by adjusting the gate voltage V in a wire networks formed by the set-up shown in Fig. (\ref{fig1}) in T-junction configuration as the building block \cite{AliceaNaturePhysics}.The exchange of Majoranas requires the motion of domain walls, driven by the current from the gate voltage \cite{Tatara2}.
Experimentally, the N\'{e}el wall with in-plane magnetization vector as shown in Fig.(\ref{fig1}) can be realized in a thin film of metallic ferromagnet  well below its Curie temperature with thickness smaller than a critical value with easy-axis anisotropy along $x$ and very large $xz$ easy-plane anisotropy. The pitch of the domain wall $\lambda=L/\nu$ depends on the exchange stiffness $A$ and easy-axis anisotropy coefficient $K$ as $\lambda\sim\sqrt{A/K}$ \cite{BookDW}. Typical values of parameters in common metallic ferromagnets such as Co, Ni and Fe are $A\simeq 10\mathrm{meV/A^o}$, $K\simeq 0.01-1\mu\mathrm{eV/{A^o}^3}$, $V_{\mathrm{ex}}\simeq 1- 10\mathrm{meV}$, which give $\lambda\simeq 100-1000\mathrm{A^o}$.The $\Delta_0\simeq 1\mathrm{meV}$ for elemental BCS superconductors such as Nb and Pb. This estimate readily puts our system to be within the MF-hosting nontrivial phase, requiring just few 10 mV's gate voltage to tune the chemical potential and drive the system across the topological phase transition.
\par
\textit{Conclusion.\textemdash}\label{sec:concl}
We propose a very simple and easy-to-build system for realization of Majorana zero modes in a magnetic domain wall in proximity with a conventional superconductor. The Majorana fermions show themselves as localized zero-energy states at the interfaces between the domain wall and ferromagnets with uniform magnetization in opposite directions. The effective spin-orbit and Zeeman terms together with the isotropic superconducting pair potential lead to an effective anisotropic $p$-wave pairing capable of harboring Majorana fermions. Our results indicate that the nontrivial phase with its MFs is achievable in realistic situations without fine tuning while the topological quantum phase transition from the trivial phase with no Majorana zero mode to the topological phase with its localized zero mode is attainable by varying the gate voltage, making the proposal very feasible practically. 

\textit{Acknowledgements.\textemdash}
We are grateful to ICTP Trieste, where this work began, for its hospitality. BZR thanks M. Zareyan, A. Nersesyan, L. Glazman, and J. Alicea for fruitful discussions. This work was partially supported by Iran Science Elites Federation (BZR) and the grants ANR-10-LABX-0037 of the Programme des Investissements d'Avenir of France (IM).

\end{document}